\documentclass[pre,showpacs,twocolumn,floatfix,bibnotes]{revtex4}
\usepackage{here}
\usepackage[dvips]{graphicx}

\newcommand\pictc[5]{\begin{figure}
                       \centerline{
                       \includegraphics[width=#1\columnwidth]{#3}}
                   \protect\caption{\protect\label{fig:#4} #5}
                    \end{figure}            }
\newcommand\pict[4][1.]{\pictc{#1}{!tb}{#2}{#3}{#4}}
\newcommand\rpict[1]{\ref{fig:#1}}
\newcommand\leqt[1]{\protect\label{eq:#1}}
\newcommand\reqtn[1]{\ref{eq:#1}}
\newcommand\reqt[1]{(\reqtn{#1})}

\newcounter{Fig}

\begin{document}

\begin{sloppy}

\title{Nonlinear guided waves and symmetry breaking in left-handed waveguides}

\author{Ilya V. Shadrivov}

\affiliation{Nonlinear Physics Group, Research School of
Physical Sciences and Engineering, Australian National
University, Canberra ACT 0200, Australia }

\begin{abstract}
We analyze nonlinear guided waves in a planar waveguide made of a left-handed material surrounded by a Kerr-like nonlinear dielectric. We predict that such a waveguide can support fast and slow symmetric and antisymmetric nonlinear modes. We study the symmetry breaking bifurcation and asymmetric modes in such a {\em symmetric} structure. We analyze nonlinear dispersion properties of the guided waves, and show that the modes can be both forward and backward.
\end{abstract}


\maketitle

Experimental realization of composite structures with simultaneously negative dielectric permittivity and negative magnetic permeability \cite{Smith:2000-4184:PRL+} revived the theoretical interest to the left-handed materials (LHMs). Such materials were first considered theoretically by Veselago \cite{Veselago:1967-2854:UFN}, who predicted negative refraction at the interface between LHM and conventional dielectric, inverse Doppler and inverse Vavilov-Cherenkov effects. Although LHMs do not exist in nature, artificial metallic composites \cite{Smith:2000-4184:PRL+} can effectively be described as materials having both dielectric permittivity and magnetic permeability negative in the microwave frequency range. Nowadays, the efforts have been made to design LHMs for optical frequencies using metallic nanowires \cite{Shalaev:2002-65:JNOM}. Additionally, it was shown that photonic crystals demonstrate negative refraction under some conditions \cite{Kosaka:1998-10096:PRB, Notomi:2000-10696:PRB}, thus resembling the most famous feature of LHMs.

In 1960's Veselago predicted that a slab of LHM with some particular parameters surrounded by usual dielectric can focus light emitted by a point source located at one side of the slab to the point image on the other side of the slab \cite{Veselago:1967-2854:UFN}. Later, Pendry suggested \cite{Pendry:2000-3966:PRL} that such a lens can focus not only propagating waves, but also evanescent waves, thus creating a perfect image of the source. Although, hot debates on the possibility to create such a perfect lens followed this suggestion (see, e.g. \cite{Garcia:2002-207403:PRL+}), it is clear that the planar structures containing LHMs have a high potential for the future image processing applications. In particular, it was already shown, that the periodic structure of alternating layers of negative refraction material and dielectric possesses an unusual band-gap structure \cite{Li:2003-083901:PRL, Shadrivov:2003-3820:APL}, when the average refractive index on the structural period is equal to zero, and it can be used for the complex beam shaping \cite{Shadrivov:2003-3820:APL}.

Recent studies of the waveguiding by the structures with LHMs have shown that the properties of guided modes in such systems differ essentially from those of conventional waveguides. So far, linear waves guided by an interface and by a slab of LHM (see, e.g. \cite{Ruppin:2000-61:PLA+, Shadrivov:2003-057602:PRE}) as well as {\em nonlinear surface waves} \cite{Shadrivov:unpub} have been studied. It has been shown that the mode structure and frequency dispersion may have an unconventional form and unexpected properties.   

The study of linear waves of a LH slab waveguide \cite{Shadrivov:2003-057602:PRE} has shown some of their peculiar properties, such as the absence of the fundamental mode, mode double degeneracy, and the existence of both forward and backward waves. In this paper, we study {\em nonlinear} guided modes in a waveguide formed by a slab of linear LHM embedded into nonlinear dielectric, and we show that symmetric, antisymmetric and asymmetric waves are supported by such a waveguide, and we study their properties. We predict, that in nonlinear regime additional modes which do not exist in linear problem appear. We demonstrate that the propagation of the wavefronts (characterized by the phase velocity) with respect to the direction of the energy flow (Poynting vector) depends on the propagation constant, and the waves can be both forward and backward travelling.

To study nonlinear guided waves in a nonlinear waveguide, we consider a LH slab with real negative dielectric permittivity $\epsilon_2$ and real negative magnetic permeability $\mu_2$ surrounded by a nonlinear dielectric [see inset in Fig. \rpict{dispers}(a)] with constant magnetic permeability $\mu_1$ and dielectric permittivity $\epsilon_1$ depending on the intensity of the electric field
\begin{equation}\leqt{Kerr_nonlinear}
\epsilon_1 \left( |E|^2 \right) = \epsilon_1 + \alpha |E|^2,
\end{equation}
where we assume $\alpha > 0$ which corresponds to the self-focusing nonlinear medium. The waveguide is uniform along $y$-axis. To be specific, we consider TE-polarized guided waves, which are governed by the nonlinear Helmholtz equation 
\begin{equation}\leqt{Helmholtz}
\Delta E_y + k_0^2 \epsilon \left(x,|E_y|^2 \right) \mu \left( x \right) E_y = 0,
\end{equation}
where $\Delta = \partial^2 / \partial x^2 + \partial^2 / \partial z^2$ is a two-dimensional Laplacian, $\epsilon \left(x,|E_y|^2 \right)$ and $\mu \left( x \right)$ are the dielectric permittivity and magnetic permeability, $k_0 = \omega/c$ is a free-space wavenumber, $\omega$ is an angular frequency, and $c$ is the speed of light. The guided waves in a similar system, but with a conventional dielectric core have been studied analytically in Ref. \cite{Akhmediev:1982-299:JETP}.

We look for stationary guided modes in the form $E = \Psi(x) \exp{(i h z)}$, where the transverse mode structure $\Psi(x)$ can be determined from the equation
\begin{equation}\leqt{mode_structure}
\frac{d^2\Psi}{dx^2} + \left[ k_0^2 \epsilon \left(x, |\Psi|^2 \right) \mu(x) 
		- h^2 \right] \Psi = 0.
\end{equation}
We introduce the dimensionless variables $(x^\prime, z^\prime) = (k_0 x, k_0 z$), $\gamma = h/k_0$, $\psi = \Psi\sqrt{\mu_1 \alpha}$. For notational simplicity we omit the primes below, and in the dimensionless variables Eq.~\reqt{mode_structure} can be rewritten in the form 
\begin{eqnarray}\leqt{mode_structure_split}
\frac{d^2 \psi}{d x^2} + \left[ \epsilon_2 \mu_2 - \gamma^2 \right] \psi 
		= 0,  \;\; |x| < L , \nonumber\\
\frac{d^2 \psi}{d x^2} + \left[ \epsilon_1 \mu_1 - \gamma^2 \right] \psi 
		+ |\psi|^2\psi = 0, \;\; |x| > L ,
\end{eqnarray}
where $2L$ is a dimensionless thickness of the slab.
\pict{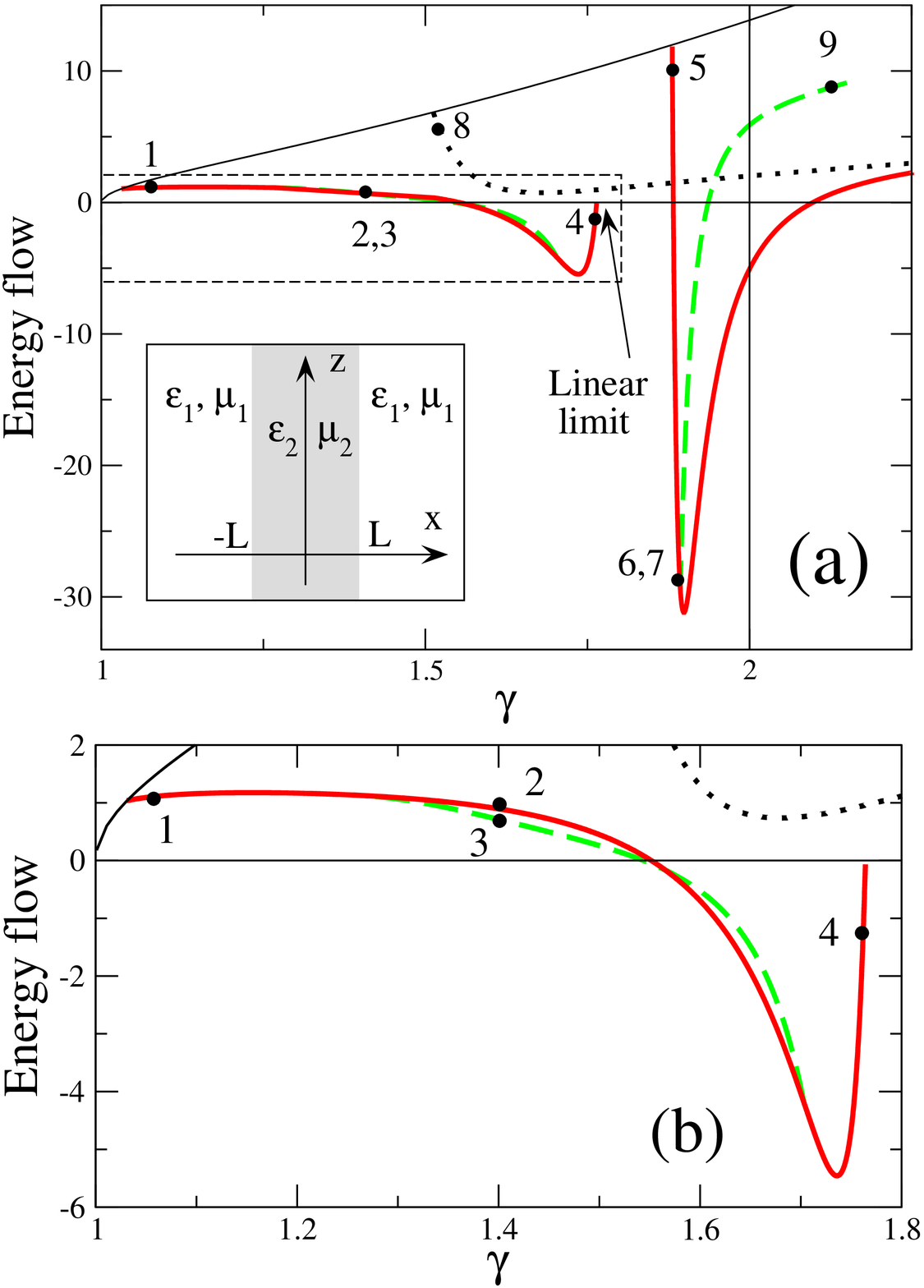}{dispers}{(color online) (a) Dependence of the normalized power of guided modes $p$ on the propagation constant $\gamma$. Parameters are: $ L = 2 $, $\epsilon_1 = 1$, $\mu_1 = 1$, $\epsilon_2 = -2$, $\mu_2 = -2$. Vertical line $\gamma = 2$ separates the fast (to the left of the line) and slow modes. Bold solid curve -- symmetric mode, dotted -- antisymmetric, dashed -- asymmetric, thin solid line -- power of two solitons in nonlinear medium vs. propagation constant. Dashed rectangle is magnified below in (b). Numbers indicate parameters for which the mode structure is shown in Fig.~\rpict{structures}.}

\pict{fig02.eps}{structures}{(color online) Mode structure calculated for the parameters indicated in Figs.~\rpict{dispers}(a,b).}

At this point, we separate the guided waves into {\em fast} and {\em slow modes}. The fast modes have the phase velocity larger than the phase velocity of light in a homogeneous medium of the core. For such modes, $\gamma^2 < \epsilon_2 \mu_2$, and their localization is caused by the total internal reflection of light from the cladding, resembling localization of waves in a dielectric waveguide. For the slow modes, $\gamma^2 < \epsilon_2 \mu_2$, and the wave guiding resembles localization of the surface waves.

Solutions of Eq.~\reqt{mode_structure_split} for guided modes can be found in the form
\begin{equation}\leqt{modes}
\psi(x) = \left\{
	\begin{array}{lr}
	\sqrt{2}\kappa_1 {\rm sech}{[\kappa_1(x-x_1)]}, \;\;& x < -L, \\
	A \sin(k_2 x) + B \cos(k_2 x), \;\; & |x|< L, \\
	\sqrt{2}\kappa_1 {\rm sech}{[\kappa_1(x-x_2)]}, \;\; &x > L, \\
	\end{array}
\right.
\end{equation}
where $\kappa_{1}^2 = \gamma^2 - \epsilon_{1} \mu_{1}$, $k_2^2 = \epsilon_{2} \mu_{2} - \gamma^2$, and $A, B, x_1, x_2$ are constants determined from continuity of the tangential components of the electric and magnetic fields at the interfaces at $x = -L$ and $x = L$. The fast modes correspond to $k_2^2>0$, while for the slow modes $k_2^2<0$. Solutions \reqt{modes} at $|x| > L$ have the form of sech-functions or solitons which are centered at $x = x_1$ and $x = x_2$ at either side of the slab. The modes with $x_1 < -L$ and $x_2 > L$ have the field maxima at the corresponding side of the waveguide. From the linear theory, it follows that solutions for the stationary modes can be found separately for the symmetric and antisymmetric modes. As has already been shown \cite{Akhmediev:1982-299:JETP}, even in a symmetric nonlinear dielectric waveguide asymmetric modes can exist. To find the asymmetric waves general solution \reqt{modes} of the differential equation \reqt{mode_structure_split} should be considered here. 
Equations for the parameters $x_1, x_2$ of the guided modes can be found from the boundary conditions by eliminating the constants $A$ and $B$,
\begin{eqnarray}\leqt{dispers_equation}
\frac{{\rm sech}(M_1) \left[\beta \tanh{(M_1)} - \tan{(k_2 L)} \right]}{  
{\rm sech}(M_2) \left[\tan{(k_2 L)} - \beta \tanh{(M_2)}\right]} = 1, 
					\nonumber\\
\frac{{\rm sech}(M_1) \left[\beta \tanh{(M_1)} + \tan^{-1}{(k_2 L)}\right]}{ 
{\rm sech}(M_2) \left[\tan^{-1}{(k_2 L)}+	\beta \tanh{(M_2)}\right]}=1,
\end{eqnarray}
where $M_1 = \kappa_1(L+x_1)$, $M_2 = \kappa_1(L-x_2)$, and $\beta = \kappa_1 \mu_2 / k_2 \mu_1$. Corresponding equations for the symmetric and antisymmetric modes (for which $-x_1 = x_2 = x_0$) can be obtained in the form: 
\begin{equation}\leqt{dispers_equation_symmetric}
\beta \tanh{[\kappa_1 (L-x_0)]} = \pm\tan^{\pm 1}{(k_2 L)},
\end{equation} 
where $(+)$ corresponds to the symmetric modes, while $(-)$ to the antisymmetric ones. Solution of Eqs.~\reqt{dispers_equation} can be found numerically after reducing them to a single equation, or the analytical solution can be obtained in the way discussed in Ref.~\cite{Akhmediev:1982-299:JETP}, where the problem has been solved separately in two different cases, $AB \ge 0$ and $AB < 0$ and the explicit expressions for unknown parameters have been obtained.

\pict{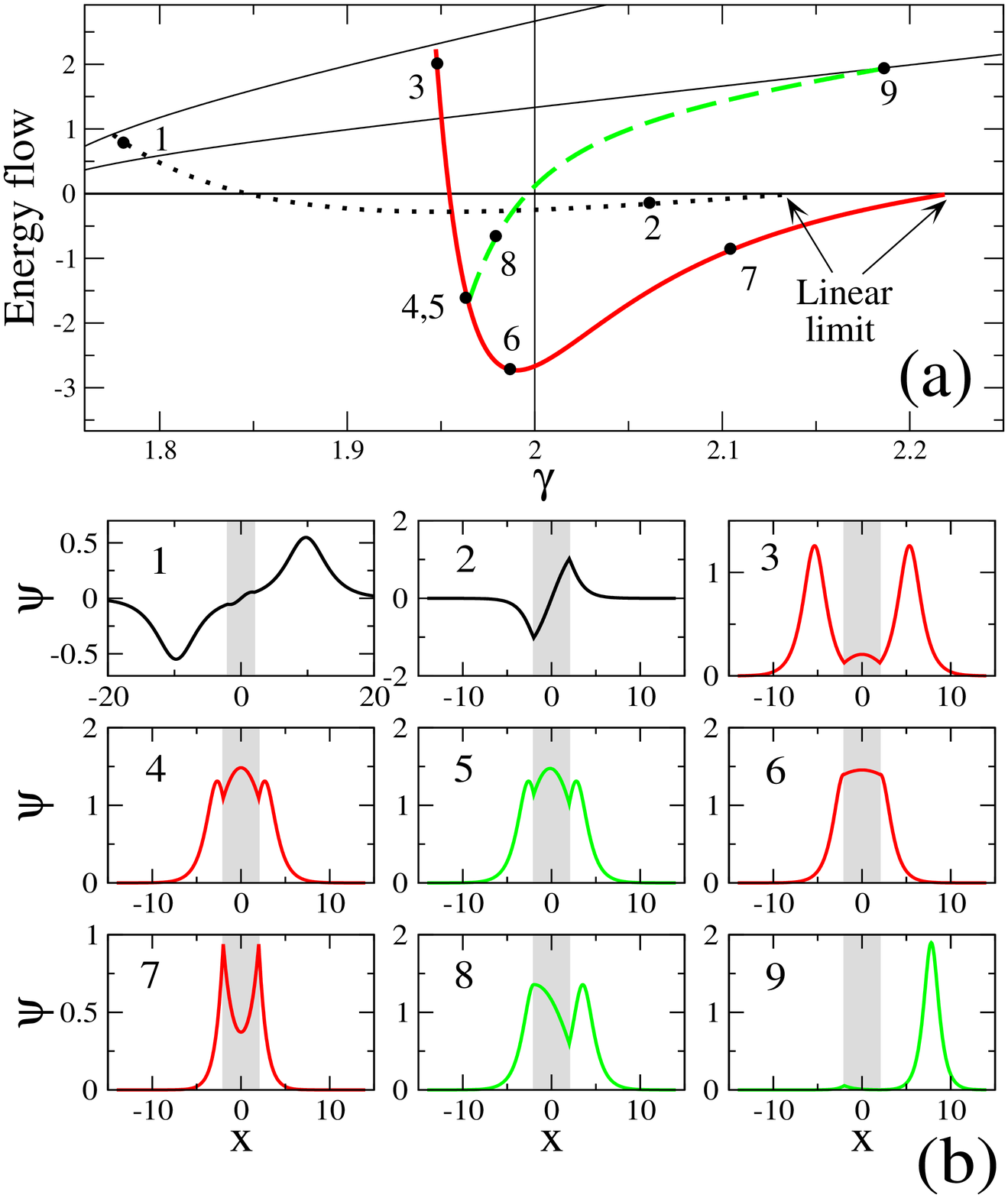}{dispers2}{(color online) (a) Dependence of the normalized power of guided modes $p$ on the propagation constant $\gamma$. Parameters are: $ L = 2 $, $\epsilon_1 = 1$, $\mu_1 = 3$, $\epsilon_2 = -2$, $\mu_2 = -2$. Vertical line $\gamma = 2$ separates the fast (to the left of the line) and slow modes. Bold solid curve -- symmetric mode, dotted -- antisymmetric, dashed -- asymmetric, two thin solid lines -- dispersion of one (lower curve) and two (upper curve) soliton states in the corresponding nonlinear media. Numbers indicate the parameters for which the mode structure is shown in (b).}

Energy flow in a stationary guided mode has the only component along the waveguide and it can be found as an integral of the Poynting vector:
\begin{equation}\leqt{energy_flow}
P = -\frac{c}{4\pi} \int_{-\infty}^{\infty}  E_y  H_x \, dx.
\end{equation} 
In the normalized form, the energy flow can be written as 
\begin{eqnarray}\leqt{energy_flow1}
p \equiv \frac{4\pi\omega}{c^2} P = \frac{\gamma\kappa_1}{\mu_1}
\left[ 2 - \tanh{(M_1)} - \tanh{(M_2)} \right] + \nonumber\\
\frac{\gamma}{2\mu_2} \left[
	L\left( B^2+A^2 \right) + \frac{\sin{(2 k_2 L)}}{2k_2}
	\left( B^2-A^2 \right)
\right].
\end{eqnarray} 
Note, that the energy flow in the mode can be both positive and negative with respect to the wavevector, since inside the slab the energy propagates in the opposite direction to that outside the slab, and the total energy flow depends on a ratio of these quantities.

Dependence of the normalized energy flow $p$ on the dimensionless wave number $\gamma$ for both fast and slow modes is shown in Fig.~\rpict{dispers}. Parameters in the figure are chosen in such a way that only one fast mode with a single node (close to the point 4) exists in the linear case (at low intensities; for the solution of linear problem and for the choice of parameters see, e.g. Ref. \cite{Shadrivov:2003-057602:PRE}). Thin solid curve shows the power of two solitons in a homogeneous nonlinear cladding versus the propagation number. The mode structures are shown in Fig.~\rpict{structures}, where each plot demonstrates the transverse wave profile corresponding to the numbered point in Fig.~\rpict{dispers}. 

The modes with the parameters close to the dotted curve (points 1, 5, 8 in Fig.~\rpict{dispers}) resembles two in-phase and out-of-phase solitons at either side of the waveguide. Closer to the dotted curve the soliton centers in the nonlinear media move further away from the waveguide. The symmetry breaking bifurcation appears on the symmetric mode branch. The asymmetric mode dispersion curve in the fast region ends in the symmetric mode branch [see Fig.~\rpict{dispers} (a)], while in the slow wave region the asymmetric mode disappears, when the amplitude of the wave at one interface becomes the same as the amplitude of the soliton on the other side of the waveguide. Two structures (points 6,7 in Fig.~\rpict{structures}) obtained at the same value of the propagation constant show the point of the symmetry breaking, where a slight asymmetry can be seen in the mode structure shown in the example 7 in Fig.~\rpict{structures}. Symmetric and asymmetric modes can be both forward and backward travelling ($p>0$ and $p<0$, respectively).

For the parameters indicated in Fig.~\rpict{dispers}, only one fast guided mode with a single node exists in the linear limit, while in the nonlinear regime the modes with zero, one and two nodes appear (see Fig.~\rpict{structures}). With increasing the waveguide thickness, one more symmetry breaking point appears on the antisymmetric mode branch, when the slab parameter $L$ exceeds some threshold value. Moreover, more high-order modes can be supported by the structure, and the nonlinear dispersion diagram becomes more complicated.

Nonlinear dispersion shown in Fig.~\rpict{dispers2}(a) is obtained for the parameters when two slow modes exist in the linear case (at low intensities). The transverse structure of the modes, corresponding to the numbered points in Fig.~\rpict{dispers2}(a) is shown in Fig.~\rpict{dispers2}(b). For symmetric and antisymmetric modes dispersion curves start from zero energy points  corresponding to linear modes in the slow-wave region of the plot, and they end at the curve representing the power of two solitons when the modes have the structure of two in-phase or out-of-phase solitons shifted from the waveguide at either side of it. Nonlinearity can only reduce the propagation constant of these modes (bold solid and dotted lines), thus increasing the wave phase velocity. These symmetric and antisymmetric modes become fast with the growth of the intensity. This becomes possible due to the nonlinear change of the index of refraction profile in the waveguide cladding. This index modification is caused by a high field amplitude in the part of the cladding next to the core, as one can see from the plots (1,3,4,6) in Fig.~\rpict{dispers2}.

The symmetry breaking bifurcation point is located at the symmetric mode branch. The asymmetric mode branch ends at the curve corresponding to the power of one-soliton, when the mode is represented by a soliton infinitely shifted from the waveguide from one side of it and a vanishingly small field trapped inside the waveguide (point 9 in Fig.~\rpict{dispers2}). Plots (4,5) in Figs.~\rpict{dispers2}(b) shows the symmetric and asymmetric modes structure for the same value of the propagation constant at the point of the symmetry breaking. 

The two presented dispersion characteristics of the nonlinear waves in a LH waveguide surrounded by a Kerr-like nonlinear medium show the general dispersion properties of the low-order fast and slow modes, which are qualitatively similar in other parameter regions.

In conclusion, we have studied nonlinear guided waves in a waveguide structure created by a slab of linear LH material surrounded by a nonlinear conventional dielectric. We have shown that fast and slow symmetric, asymmetric and antisymmetric modes can exist in this system. On the dispersion diagram, the symmetry breaking occurs for both symmetric and antisymmetric mode branches. In the latter case, there exists a threshold value of the slab thickness above which the asymmetric modes appear. The guided waves can be both forward and backward travelling. While the direction of the energy flow is determined by a source, the type of the wave determines the phase velocity direction, i.e. the phase front propagates from the source in the forward waves, and to the source, in the backward waves. 

The author thanks Yuri S. Kivshar and Andrey A. Sukhorukov for useful discussions. This work was partially supported by the Australian Research Council.

\end{sloppy}
\end{document}